\documentclass[aps,prl,twocolumn,superscriptaddress,preprintnumbers]{revtex4-1}
\pdfoutput=1
\usepackage{amsmath}
\usepackage{graphicx}
\usepackage{bbm}
\usepackage{amssymb}
\usepackage{enumitem}
\usepackage{booktabs}
\usepackage{verbatim}
\usepackage{soul}
\usepackage{mathrsfs}
\usepackage[linktocpage=true,
colorlinks=true,
urlcolor=blue,
anchorcolor=blue,
citecolor=blue,
filecolor=blue,
linkcolor=blue,
menucolor=blue,
]{hyperref}

\newcommand{\ex}{\mathrm{e}}
\newcommand{\diff}{\mathrm{d}}
\newcommand{\dd}{\mathrm{d}}
\newcommand{\R}{\mathbb{R}}
\newcommand{\vol}{\mathrm{vol}}

\newcommand{\C}{\mathbb{C}}

\newcommand{\hook}{\mathbin{\rule[.2ex]{.4em}{.03em}\rule[.2ex]{.03em}{.9ex}}}

\usepackage{color}

\def\nn{\nonumber}
\newcommand{\ii}{\mathrm{i}}

\newcommand{\Z}{\mathbb{Z}}

\begin{document}

\title{Localizing Wrapped M5-branes\\ and Gravitational Blocks}

\author{Pietro Benetti Genolini}
\affiliation{Department of Mathematics,
King's College London, Strand, London, WC2R 2LS, U.K.}
\author{Jerome P. Gauntlett}
\affiliation{Blackett Laboratory, Imperial College, Prince Consort Road, London, SW7 2AZ, U.K.}
\author{James Sparks}
\affiliation{Mathematical Institute, University of Oxford, Woodstock Road, Oxford, OX2 6GG, U.K.}

\begin{abstract}
\noindent  We consider $d=2$, $\mathcal{N}=(0,2)$ SCFTs that can arise from M5-branes 
wrapping four-dimensional, complex, toric manifolds and orbifolds. We use equivariant localization to compute the off-shell central charge of the dual supergravity solutions, obtaining a result which can be written as a sum of gravitational blocks and precisely agrees with a field theory computation using anomaly polynomials and $c$-extremization.
\end{abstract}

\maketitle

\section{Introduction}\label{sec:intro}

Supersymmetric wrapped branes continue to provide a fertile arena for exploring the AdS/CFT correspondence. They give rise to rich classes of novel SCFTs in various spacetime dimensions, and they also provide a concrete framework for obtaining a microstate interpretation of the Bekenstein--Hawking entropy for asymptotically AdS black holes. In addition, the supersymmetric AdS solutions of supergravity associated with wrapped branes give rise to novel geometric structures which are of interest in their own right.

In a recent paper \cite{BenettiGenolini:2023kxp}, a new calculus was introduced for supersymmetric solutions of supergravity that have an R-symmetry.
For several general classes of such solutions, it was shown there exists a set of equivariantly closed
differential forms which can be constructed from Killing spinor bilinears. Furthermore, various BPS observables can then be computed
using localization via the Berline--Vergne--Atiyah--Bott (BVAB) fixed point formula \cite{BV:1982, Atiyah:1984px}, without solving the supergravity equations of motion. Here we want to further develop
these tools for a general class of $AdS_3$ solutions of $D=11$ supergravity that arise from M5-branes wrapping four-dimensional manifolds and orbifolds. 
The preserved supersymmetry is such that the $AdS_3$ solutions are dual to $d=2$, $\mathcal{N}=(0,2)$ SCFTs and, in particular, they have 
an R-symmetry. 

More precisely, we focus on the class of supersymmetric $AdS_3\times M_8$ solutions of $D=11$ supergravity considered in \cite{Gauntlett:2006qw} and further analysed in \cite{Ashmore:2022ydf}.
We construct a set of equivariantly closed forms and show that they can be used to compute the central charge of the dual SCFT, as well as the conformal dimensions of operators in the SCFT that are dual to supersymmetric wrapped probe M2-branes. 
To illustrate the formalism, we focus on examples where $M_8$ is an $S^4$ fibration over a toric $B_4$, which 
are associated with M5-branes wrapping $B_4$. Focusing on toric $B_4$ is of interest since we 
can both compare with some known and conjectured field theory results, as well as obtain results which provide new 
field theory predictions. 
As we shall see the localization results are remarkably simple for these examples because the fixed points of the R-symmetry, which is 
linear combination of the $U(1)^2$ action on $S^4$ and the $U(1)^2$ action on $B_4$,
are a set of isolated points on $M_8$. We can use the BVAB formula to implement flux quantization as well as obtain an 
\emph{off-shell} expression for the central charge. After extremizing over the undetermined R-symmetry data, 
we then obtain an on-shell expression for the central charge. As explained in more detail in \cite{BenettiGenolini:2023ndb}, it is important to emphasize that this will give the correct 
central charge, without solving the supergravity equations, just assuming the supergravity solution actually exists, or equivalently,
that the low energy limit of M5-branes wrapped on the specific toric $B_4$ does indeed flow to a SCFT in the IR, in the large $N$ limit.
The formalism therefore provides a geometric, off-shell version of $c$-extremization \cite{Benini:2012cz, Benini:2013cda}.

The off-shell expression for the central charge that we derive takes the form of a sum of gravitational blocks \cite{Hosseini:2019iad}.
The terminology `gravitational block'  is perhaps somewhat confusing, as in many works it is not referring to a computation in gravity, but rather is a conjecture that should arise for some unspecified gravity computation.
More precisely various off-shell expressions for BPS quantities have been proposed which either can be derived in field theory, for example using anomaly polynomials, or alternatively have been noted to give  the correct on-shell result for some specific, explicitly known supergravity solutions. For the former, invoking AdS/CFT, there is then an expectation that there will be a corresponding off-shell computation within gravity that leads to the same off-shell result for the central charge. However, it is not at all clear, in general, how one should go off-shell on the gravity side. That being said, 
in the setting of GK geometry, for Sasaki--Einstein fibrations over spindles, this was recently achieved in \cite{Boido:2022mbe}.  The results of this paper, as well
as \cite{BenettiGenolini:2023kxp,BenettiGenolini:2023ndb} indicate that the equivariant calculus of \cite{BenettiGenolini:2023kxp} provides a universal way of deriving gravitational blocks within a gravitational context. 
Moreover, the new results make it clear that the origin of gravitational blocks is when a trial R-symmetry has isolated 
fixed points on the space that the brane is wrapping.

In \cite{BenettiGenolini:2023ndb} we provide some further details of the equivariant calculus for the general class of $AdS_3$ solutions
of $D=11$ supergravity discussed here. In addition, we will also analyse other examples of wrapped M5-branes, where the R-symmetry fixed point set no longer consists of isolated points and, in particular, gravitational blocks are not relevant. 

\section{\texorpdfstring{$AdS_3\times M_8$}{AdS3 x M8} Solutions}\label{sec:M8}

\newcommand{\DA}{\lambda}
\newcommand{\chiA}{\epsilon}
\newcommand{\LA}{\xi}
\newcommand{\PhiA}{\Psi}

We consider supersymmetric solutions of $D=11$ supergravity of the form
\begin{align}
\diff s^2&= \ex^{2\lambda}\left[ \dd s^2(AdS_3)+ \dd s^2(M_8)\right]\,,\nn\\
G&=\ex^{3\lambda}F + \vol(AdS_3)\wedge \ex^{3\lambda}f\,,
\end{align}
where $\lambda,F$ and $f$ are a function, a four-form and a one-form on $M_8$,
respectively.
In addition, $\dd s^2(AdS_3)$ is the metric on a unit radius $AdS_3$ 
and $\vol(AdS_3)$ is the corresponding volume form. 
The Bianchi identity implies $\dd(\ex^{3\lambda} f)=0$, and it is convenient to introduce a function $a_0$, locally defined in general, 
via $\ex^{3\lambda} f= \dd a_0$.

We assume that the preserved supersymmetry is such that the dual $d=2$ SCFTs have $\mathcal{N}=(0,2)$ supersymmetry.
We will focus on the class of solutions classified in~\cite{Gauntlett:2006qw}: following the conventions of \cite{Ashmore:2022ydf}, there is then
a complex spinor $\epsilon$ on $M_8$, with $\bar\epsilon \epsilon=1=\bar \epsilon^c\epsilon$ as well as $\bar\epsilon^c\gamma_9\epsilon=0$, where
$\gamma_9\equiv \gamma_1\dots \gamma_8$. There is an R-symmetry Killing vector $\xi$, with a dual one-form $\xi^\flat$ which can be constructed 
as a bilinear:
\begin{align}
\xi^\flat =-\tfrac{\ii}{2}\bar\epsilon\gamma_9\gamma_{(1)}\epsilon\,.
\end{align}
We have introduced the notation $\gamma_{(r)}=\tfrac{1}{r!}\gamma_{\mu_1\cdots \mu_r}\diff x^{\mu_1}\wedge\cdots\wedge\diff x^{\mu_r}$, and 
have 
normalized $\xi$ so that $\mathcal{L}_\xi \epsilon=\tfrac{\ii}{2}\epsilon$.
We also define a scalar, two two-forms and a four-form bilinear
\begin{align}
\sin\alpha & = \bar\chiA\gamma_9\chiA \, ,\qquad \quad \  \ \, 
 J  =-\ii\bar\chiA\gamma_{(2)}\chiA\,, \nonumber\\
 \omega & = -\ii\bar\chiA{\gamma_9}\gamma_{(2)}\chiA \, , \quad  \Psi  = 
\bar\chiA \gamma_{(4)}\chiA\, , 
\end{align}
and introduce the locally defined function $y$, given by $y=\tfrac{1}{2}(\ex^{3\lambda}\sin\alpha-a_0)$.

These ingredients can be used to define the following polyforms on $M_8$
\begin{align}\label{ads3fluxeqform2}
\Phi 
&= \ex^{9\lambda}\, \vol_8 + \tfrac{1}{4} \ex^{9\lambda} * J - \tfrac{1}{8}y\,  \ex^{6\lambda}\PhiA - \tfrac{1}{16} y^2 \ex^{3\lambda}F \nn \\
& \ \ \ + \tfrac{1}{32}y^2 \ex^{3\lambda}\omega + \tfrac{1}{192 }y^3\,,\nn\\
\Phi^F
&= \ex^{3\lambda} F-\tfrac{1}{2}\ex^{3\lambda}\omega-\tfrac{1}{4} y\,,\nn\\
\Phi^{*F}
 &= \ex^{6\lambda}* F - a_0 \ex^{3\lambda} F\nn \\
 & \ \ \ -\tfrac{1}{2} \left( \ex^{6\lambda}J - a_0 \ex^{3\lambda}\omega \right) -\tfrac{1}{4}y^2\, ,
\end{align}
where $*$ denotes the Hodge dual and $\vol_8$ is the volume form on $M_8$. 
A key result \footnote{See \cite{BenettiGenolini:2023ndb} for more details.}
is that the differential and algebraic conditions satisfied by the above bilinears, along with the Bianchi identity and equation of
motion for the four-form, imply these polyforms
are equivariantly closed: $\dd_\xi \Phi =\dd_\xi \Phi^F=\dd_\xi \Phi^{*F}=0$, where $\diff_\xi\equiv \diff-\xi\hook\, $. Thus, we can compute their integrals on closed cycles using the BVAB formula. In particular, the integral of $\Phi^F$ on a four-cycle $\Gamma_4$ represents the flux of the four-form of eleven-dimensional supergravity, which (in the large $N$ limit) should be quantized as
\begin{equation}\label{nfluxcond}
	N_{\Gamma_4} \equiv \frac{1}{(2\pi\ell_p)^3} \int_{\Gamma_4} \Phi^F \in\mathbb{Z}  \, , 
\end{equation}
where $\ell_p$ is the Planck length. 
By computing the effective three-dimensional Newton constant, one can show that
the integral of $\Phi$ is proportional to the trial central charge 
\begin{equation}
	c = \frac{3}{2^5 \pi^7 \ell_p^9} \int_{M_8} \Phi \, .
\end{equation}

\section{M5-branes wrapped on \texorpdfstring{$B_4$}{B4}}

Within the above setup, we are interested in solutions that describe holographic duals to M5-branes wrapping a holomorphic four-cycle $B_4$ inside a Calabi--Yau four-fold.
A local model for the Calabi--Yau is given by the sum of two line bundles $\mathcal{L}_1 \oplus \mathcal{L}_2 \to B_4$ subject to the condition
\begin{equation}\label{CYcond}
	c_1(\mathcal{L}_1) + c_1(\mathcal{L}_2) + c_1(TB_4) = 0 \, ,
\end{equation}
which also guarantees the supersymmetry of a wrapped M5-brane. For the associated supergravity solutions (in the near horizon limit),
we take $M_8$ to be an $S^4$ bundle over $B_4$, 
\begin{align}
S^4\hookrightarrow M_8\to B_4\, .
\end{align}
Here we write $S^4 \subset \mathbb{C}_1 \oplus \mathbb{C}_2 \oplus \mathbb{R}$,  where the $\mathbb{C}_i$ factors are twisted by the respective line bundles $\mathcal{L}_i$. 
We will assume that the solutions have $U(1)^2 \subset SO(5)$ isometry of the $S^4$, as well as the isometries of $B_4$.
In the following, we will first consider the $B_4$ base to be a complex toric surface
and compute the corresponding trial central charge using equivariant localization.
Later we will consider cases when the toric $B_4$ has orbifold singularities and we will then also slightly
 generalize the Calabi-Yau condition \eqref{CYcond}. Other classes of $B_4$ are considered in \cite{BenettiGenolini:2023ndb}.

For the toric $B_4$ examples considered here, the R-symmetry will only have isolated fixed points
and as a consequence the BVAB formula takes a particularly simple form. On $M_8$, and even-dimensional invariant submanifolds $M_{2k}\subset M_8$, the integral of a general equivariantly closed polyform {$\Phi$} is given by a sum of contributions from the fixed points $x_\ell$
\begin{align}
\label{Philocalize}
	\int_{M_{2k}} \Phi = \sum_{\ell} \frac{1}{d_\ell} \frac{(2\pi)^k}{ \epsilon_1^\ell \cdots \epsilon_k^\ell } \Phi \rvert_{x_\ell} \, .
\end{align}
Here $M$ can have orbifold singularities, where
the normal space to the point $x_\ell$ is $\R^{2k}/\Gamma_\ell$ and $d_\ell$ is the order of the finite group $\Gamma_\ell$. On this normal space $\xi$ generates a linear isometric action with weights $\epsilon^\ell_i$.

In the sequel, for simplicity, we impose one other condition on the class of solutions that we are considering; namely, that a certain flux integral threading the $S^4$ vanishes \footnote{This condition holds for the known solutions  in 
\cite{Gauntlett:2006qw, Benini:2013cda} with $M_8$ an $S^4$ fibration over $B_4$. One can relax this condition and obtain off-shell expressions for the central charge that will depend on this flux. It would be interesting to know whether or not any such solutions exist.}:
\begin{align}\label{funnyflux}
\int_{S^4}(\ex^{6\lambda}* F - a_0 \ex^{3\lambda} F)=0\,.
\end{align}

\section{Smooth toric base}

The first family of solutions we focus on is when the base $B_4$ is a toric complex surface, with $B_4$ having $U(1)^2$ isometry. 
The R-symmetry Killing vector $\xi$ on $M_8$ generically mixes the $U(1)^2 \subset SO(5)$ isometry of the $S^4$ with the $U(1)^2$ of $B_4$, 
and so we can write
\begin{equation}\label{xi}
	\xi = \sum_{i=1}^2 b_i \partial_{\varphi_i} + \sum_{A=1}^2 \varepsilon_A \partial_{\psi_A} \, ,
\end{equation}
where $b_i$, $\varepsilon_A$ are constants.
Here $\partial_{\varphi_i}$ rotate the two copies of $\mathbb{C}_i$ in $S^4 \subset \mathbb{C}_1 \oplus \mathbb{C}_2 \oplus \mathbb{R}$, 
with weight 1, and $\partial_{\psi_A}$ are a lift of the generators of the torus isometry of $B_4$ to $M_8$. 

For generic $b_i,\varepsilon_A$,  the fixed points of the action of $\xi$ on $M_8$, where $\lVert \xi \rVert = 0$, are isolated, as noted above.
Concretely, the $U(1)^2$ action on the $S^4$ has two fixed points, at the north and south pole. If we take the $U(1)^2$ isometry on $B_4$ to have $d$ isolated fixed points we then
have  a total of $2 d$ fixed points on $M_8$. These are labelled by $(N/S,a)$, where $N/S$ refers to the north or south pole of $S^4$, and $a=1, \dots, d$ labels the isolated fixed points on~$B_4$.

We can now use the BVAB formula to compute the flux of $\Phi^F$ through the $S^4$ cycle over any of the $d$ fixed points on the base. 
Since these cycles are all in the same homology class \footnote{Take any curve connecting two fixed points $x_a$, $x_{a'}$ in $B_4$, together with the $S^4$ bundle over that curve. This is a five-dimensional submanifold of $M_8$ with boundaries $S^4_a$ and $S^4_{a'}$ (the spheres over the fixed points), showing the latter are homologous.},
using
\eqref{nfluxcond} we have
\begin{align}\label{NS4}
	N_{S^4} & = \frac{1}{(2\pi \ell_p)^3} \int_{S^4} \Phi^F \nonumber\\
& = - \frac{1}{(2\pi \ell_p)^3} \frac{1}{4} \frac{(2\pi)^2}{b_1^a b_2^a} \left( y_N^a - y_S^a \right) \, ,
\end{align}
where $N_{S^4}$ is the number of wrapped M5-branes.
Here  $y_{N/S}^a$ denotes the value of the function $y$ at the fixed point $(N/S,a)$, and $b_i^a$ are the weights of the action of $\xi$ on the normal space $\R^4=\C\oplus\C$ to the fixed point in $S^4$. We can similarly compute the flux of $\Phi^{*F}$ through the same cycles which, recall from \eqref{funnyflux}, we assumed to vanish. Utilizing the BVAB
formula then immediately gives $(y_N^a)^2 = (y_S^a)^2$.
Thus, requiring that $N_{S^4} \neq 0$ we conclude that
\begin{align}
\label{eq:ya}
y_N^a = - y_S^a = - 4 \pi \ell_p^3 b_1^a b_2^a N_{S^4}\,.
\end{align}

We can similarly evaluate the central charge, with contributions from the $2d$ fixed points given by
\begin{align}\label{centralchge1}
	c &
	= \frac{3}{2^5\pi^7 \ell_p^9} \frac{1}{192} \sum_{a=1}^d \frac{(2\pi)^4}{\epsilon_1^a \epsilon_2^a b_1^a b_2^a} \left[ (y_N^a)^3 - (y_S^a)^3 \right]\nonumber \\
	&=   \sum_{a=1}^d \frac{1}{\epsilon_1^a \epsilon_2^a} (b_1^ab_2^a)^2 (-N_{S^4})^3\,.
\end{align}
Here the normal space to the fixed points in $M_8$ is $\R^8=\C^{\oplus 4}$, with $b_i^a$, $\epsilon_A^a$
being the associated weights of the action of $\xi$ on those four copies of $\C$.

It is remarkable how simply the key expression \eqref{centralchge1}, as a sum of blocks, emerges from our formalism. In particular, we see that each block is related to the off-shell central charge for the 
$d=6$, $\mathcal{N}=(0,2)$ SCFT in the large $N$ limit \footnote{Recall that this is given by 
$a=\frac{256}{7} (b_1b_2)^2N^3$, subject to $b_1+b_2=1$. This off-shell expression only assumes the existence of a $U(1)^2\subset SO(5)$ R-symmetry and after extremizing $a$ over the constrained $b_i$, one finds the on-shell weights $b_1=b_2=\frac{1}{2}$ and on-shell central charge $a=\frac{16}{7}N^3$, which is the correct result for 
the $d=6$, $\mathcal{N}=(0,2)$ SCFT with its $SO(5)$ R-symmetry, in the large $N$ limit. 
Further discussion in the context of equivariant localization can be found in appendix B of \cite{BenettiGenolini:2023ndb}.}. 
To obtain our final off-shell result it remains
to compute $b_i^a$, $\epsilon_A^a$ in terms of the R-symmetry vector~\eqref{xi}, together with global topological data for $M_8$.
 In fact we will be able to do this straightforwardly,  utilizing various standard results in the toric geometry literature.

\section{Weights from toric geometry}\label{sec:weights}

We begin by recalling some key facts about complex toric four-manifolds $B_4$. 
By definition these are complex manifolds equipped with a holomorphic 
$(\C^*)^2$ action, which has a dense open orbit. There always exists a compatible K\"ahler metric, where $U(1)^2\subset (\C^*)^2$ is an isometry, but we emphasize that no metric data 
enters the fixed point formulae we  use.  

Such a $B_4$ has a distinguished set of  $a=1,\ldots,d$ toric divisors $D_a\subset B_4$. 
By definition these are complex two-dimensional submanifolds, invariant  under the $U(1)^2$ action, where the 
normal space to a given $D_a$ is  rotated by the $U(1)\subset U(1)^2$ subgroup specified by 
a vector with components $v^a_A\in \Z$, $A=1,2$. The 
set of $v^a_A$ is referred to as the toric data for $B_4$.
The toric divisors may be 
 ordered 
cyclically, with $x_a=D_a\cap D_{a+1}$ precisely giving the 
set of $d$ points that are fixed under the $U(1)^2$ action, the index 
$a$ understood to be defined modulo $d$. 

If $\partial_{\psi_A}$ denote vector fields generating the 
$U(1)^2$ isometry, then we may write a  Killing vector on 
$B_4$ as $\sum_{A=1}^2\varepsilon_A \partial_{\psi_A}$. 
The  fixed point $x_a$ has normal space $\R_a^4=\C^a_1\oplus\C^a_2$, and the 
weights  $\epsilon_A^a$ of this Killing vector on $\C_A^a$  are given by the standard toric geometry formulae
\begin{align}\label{epsilon12}
\epsilon_1^a = -\det (v^{a+1},\varepsilon)\, , \quad \epsilon_2^a = 
\det (v^a,\varepsilon)\, .
\end{align}

The internal space $M_8$ is in turn the total space of an $S^4$ bundle 
over $B_4$. By definition the vector fields $\partial_{\varphi_i}$ in \eqref{xi} rotate the two copies of $\mathbb{C}_i$ in $S^4 \subset \mathbb{C}_1 \oplus \mathbb{C}_2 \oplus \mathbb{R}$, 
with weight 1, but to define $\xi$ we must also choose a lifting of the 
$\partial_{\psi_A}$ to $M_8$. This may be achieved by 
choosing a lifting to each line bundle $\mathcal{L}_i\rightarrow B_4$, 
making these equivariant line bundles. On the other hand, a basis 
of such equivariant line bundles $L_a$ is naturally provided by 
the toric divisors $D_a$. The corresponding equivariant first Chern class 
$c_1^{\xi}(L_a)$,  when restricted to the fixed point $x_b\in B_4$, 
is given by the formula
\begin{align}\label{c1La}
c_1^{\xi}(L_a)\mid_{x_b} = \delta_{a,b}\epsilon_1^a + \delta_{a, b+1}\epsilon_2^b\, ,
\end{align}
where the weights $\epsilon^a_A$ are given by \eqref{epsilon12}. 
We may thus write 
$\mathcal{L}_i= -\sum_{a=1}^d \mathfrak{p}_i^aL_a$, 
where  $\mathfrak{p}_i^a\in \Z$ specify both the topology of $\mathcal{L}_1\oplus\mathcal{L}_2
\rightarrow B_4$, and also a choice of lifting of the $U(1)^2$ isometry of 
$B_4$ to the total space. From \eqref{c1La} the weights of $\xi$ on
the two complex line fibres are then
\begin{align}\label{bi}
b_i^a = b_i - \sum_{b=1}^d \mathfrak{p}^b_i c_1^{\xi}(L_b)\mid_{x_a} 
= b_i - \mathfrak{p}^a_i\epsilon_1^a - \mathfrak{p}^{a+1}_i \epsilon_2^a\, .
\end{align}

Having determined explicit formulae \eqref{epsilon12}, \eqref{bi}
for the weights of $\xi$ at the fixed points, finally we must impose that 
$\xi$ is an R-symmetry: that is, there is a Killing spinor $\epsilon$  satisfying
 $\mathcal{L}_\xi \epsilon=\tfrac{\ii}{2}\epsilon$. 
This is where the Calabi--Yau condition \eqref{CYcond} enters 
as a further set of constraints on our parameters. 

For a toric complex 
manifold $B_4$ 
we have the standard toric geometry formula
$c_1(TB_4) = \sum_{a=1}^d c_1(L_a)$. 
Since also by definition $\mathcal{L}_i= -\sum_{a=1}^d \mathfrak{p}_i^aL_a$, 
imposing the equivariant version of equation \eqref{CYcond} gives
\begin{align}\label{c1sum}
\sum_{a=1}^d \left(\sum_{i=1}^2 \mathfrak{p}_i^a -1\right) c_1^\xi(L_a) = 0 \, .
\end{align}
The $c_1^\xi (L_a)$ are precisely a set of generators for the equivariant cohomology 
of $B_4$, with no relations, so the coefficients in \eqref{c1sum} must all be zero: $\sum_{i=1}^2 \mathfrak{p}_i^a =1$.

On the other hand, the resulting $SU(4)$-invariant chiral spinor on the Calabi--Yau 
four-fold satisfies a standard set of projection 
conditions $\gamma^{2j-1,2j}\epsilon = \ii \epsilon$ in an orthonormal frame, for 
each $j=1,2,3,4$. 
The original local Calabi--Yau geometry 
is embedded inside $M_8$ as the normal bundle of the north pole 
section, in our conventions for the labelling of north/south poles. 
 As shown in the appendix, given the above projection conditions the charge  of the spinor 
at the point $x_a^N$ in this north pole section is then
\begin{align}
\mathcal{L}_\xi \epsilon = \frac{\ii}{2}(b_1^a + b_2^a + \epsilon^a_1 + \epsilon^a_2 ) \epsilon
= \frac{\ii}{2}(b_1+b_2) \epsilon \, .
\end{align}
Here we have used \eqref{bi}, and then imposed the Calabi--Yau condition in the form \eqref{c1sum}. Thus,
together we have the following constraints on our parameters:
\begin{align}\label{cons}
b_1+b_2=1\, , \qquad \sum_{i=1}^2 \mathfrak{p}_i^a =1\, .
\end{align}
We will discuss a generalization of these constraints later. 

Inserting the formulae \eqref{epsilon12}, \eqref{bi} into 
\eqref{centralchge1} gives our final `gravitational block' formula for the trial central 
charge in supergravity, expressed in terms of the toric data $v^a_A$, 
$\mathfrak{p}^a_i$, and choice of R-symmetry vector field \eqref{xi}, subject
to \eqref{cons}.
Our resulting expression agrees precisely with the field theory 
formula in \cite{Hosseini:2020vgl}, obtained 
via equivariant localization of the M5-brane anomaly polynomial,  and provides
a striking confirmation of AdS/CFT. 
To get the on-shell result, we need to extremize over the choice of R-symmetry; in 
\cite{BenettiGenolini:2023ndb} we show that on the gravity side this is indeed a necessary condition for 
imposing the supergravity equations of motion.

\section{Other observables}\label{sec:observables}

As also shown in \cite{BenettiGenolini:2023kxp,BenettiGenolini:2023ndb}, other 
physical observables may similarly be computed using 
equivariant localization in supergravity. 

For example, 
consider the four-cycles $\Gamma^a_i\subset M_8$ 
that are the total spaces of $S^2_i$ bundles 
over the toric divisors $D_a\subset B_4$, where 
$S^2_i\subset \C_i\oplus\R$ is a linearly embedded two-sphere 
within the fibre $S^4$. There are four fixed points, namely the copies of $x_a$, $x_{a-1}$ 
at the north and south poles of the fibre $S^2_i$. In fact $x_a$, $x_{a-1}$ 
are also precisely the poles of $D_a\cong S^2_a$, with  $\epsilon_2^a$ and $\epsilon_1^{a-1}=-\epsilon^a_2$ being 
the weights on the tangent space, respectively. 
Picking for instance $i=1$, using localization we compute the flux
\begin{align}
N^a_1 &= \frac{1}{(2\pi \ell_p)^3}\int_{\Gamma^a_1}\Phi^F  \nn \\
&= -\frac{1}{(2\pi \ell_p)^3}\frac{(2\pi)^2}{4} \left[\frac{y_N^a - y_S^a}{b_1^a \epsilon_2^a} + \frac{y_N^{a-1} - y_S^{a-1}}{b_1^{a-1} \epsilon_1^{a-1}}\right] \nn \\
&= \frac{1}{\epsilon^a_2} \left[ \mathfrak{p}_2^a ( \epsilon^{a+1}_2 + \epsilon^{a-1}_2) - (\mathfrak{p}_2^{a+1} + \mathfrak{p}_2^{a-1}) \epsilon_2^a \right] N_{S^4} \nn \\
\label{eq:Nai}
&= - \mathfrak{q}^a_2 \, N_{S^4} \, ,
\end{align}
and similarly $N^a_2= - \mathfrak{q}^a_1 \, N_{S^4}$.
To get the third line in \eqref{eq:Nai} we substituted \eqref{eq:ya}, \eqref{bi} and used \eqref{epsilon12}. 
To get the final line we evaluated the determinants in \eqref{epsilon12}, subject
to $\det(v^{a-1}, v^a)=\det(v^{a}, v^{a+1})=1$, which 
follow from $B_4$ being smooth. In \eqref{eq:Nai}
$\mathfrak{q}_i^a\in \Z$ is (minus) the integral of the first Chern class of $\mathcal{L}_i$ through the divisor $D_a$
\begin{equation}
	\mathfrak{q}_i^a \equiv - \int_{D_a} c_1(\mathcal{L}_i) = \sum_{a,b} D_{ab}\,  \mathfrak{p}^b_i \, ,
\end{equation}
where $D_{ab}=\int_{B_4}c_1(L_a)c_1(L_b)$ is the intersection matrix of the toric divisors
\begin{equation}
	D_{ab} = \begin{cases} 1 & b = a \pm 1 \\ - \det( v^{a-1}, v^{a+1}) & b = a \\ 0 & \text{otherwise} \end{cases} \, .
\end{equation}
A derivation of \eqref{eq:Nai}, using only algebraic topology, can be found in an appendix to \cite{BenettiGenolini:2023ndb}.

We can also compute the dimension of chiral primary operators dual to M2-branes wrapping submanifolds calibrated by $\omega$. These are obtained by applying the BVAB formula to $\Phi^F$ restricted to $\Sigma_2$ \cite{BenettiGenolini:2023ndb}. 
Consider the (homologically trivial) $S^2_i$ just considered over a fixed point $x_a$; if it is calibrated by $\omega$ then
\begin{align}
	\Delta(S^2_i) &= \frac{1}{(2\pi)^2\ell_p^3} \int_{S^2_i} \ex^{3\lambda} \omega = \frac{1}{(2\pi)^2\ell_p^3} \frac{2\pi}{b_i^a} \frac{ y_N^a - y_S^a }{2} \nn \\
	\label{DeltaS2}
	&= \frac{2 b_1^a b_2^a}{b_i^a} (-N_{S^4}) \, .
\end{align}
One can also similarly consider the divisor $D_a$ at, say, the north pole of the $S^4$, and if this is calibrated by $\omega$ then (with an appropriate orientation choice)
\begin{align}
	\Delta(D_a) &= - \frac{1}{(2\pi)^2\ell_p^3} \int_{D_a} \ex^{3\lambda} \omega \nn\\
& = - \frac{1}{(2\pi)^2\ell_p^3} \frac{2\pi}{2} \left( \frac{y_{N}^a}{\epsilon^2_a} + \frac{y_{N}^{a-1}}{\epsilon^1_{a-1}} \right) \nn \\
	&=  \frac{1}{\epsilon^2_a} \left(  b_1^a b_2^a - b_1^{a-1} b_2^{a-1} \right) N_{S^4} \nn \\
	\label{eq:DeltaDa}
	&=  (b_1 \mathfrak{q}_2^a + b_2 \mathfrak{q}_1^a + D_{abc} \mathfrak{p}^b_1 \mathfrak{p}^c_2 ) (-N_{S^4})\, .
\end{align}
The same expression holds for the divisor $D_a$ at the south pole of $S^4$, up 
to an overall sign related to the choice of orientation.
Here we introduced the intersection of three equivariant Chern classes (see e.g. \cite{Martelli:2023oqk} for a review)
\begin{align}\label{Dabc}
	& D_{abc} \equiv \int_{B_4} c_1^\xi(L_a) c_1^\xi(L_b) c_1^\xi( L_c) \nn\\
	&= \begin{cases} - \epsilon_2^{a-1} & a_i = a_j = a_k+1 \equiv a \\ - \epsilon_1^a & a_i = a_j = a_k-1 \equiv a \\ - (\epsilon_1^a)^2 - (\epsilon_2^{a-1})^2 &  a_i = a_j = a_k \equiv a \\ 0 & \text{otherwise} \end{cases} \, .
\end{align}
Notice that the first two terms in the last line of \eqref{eq:DeltaDa}, as well as the expression \eqref{DeltaS2}, are both independent of $\varepsilon_A$, and indeed match the analogous formulas in the absence of $U(1)^2$ symmetry on $B_4$ discussed in \cite{BenettiGenolini:2023ndb}. In contrast, this is not true of $D_{abc}$ in \eqref{eq:DeltaDa}, and in fact $\Delta(D_a)$ is linear in $\epsilon^a_A$.

\section{Orbifolds and anti-twists}

The above discussion has some immediate generalizations,  and this allows us to make a connection with the results of \cite{Faedo:2022rqx, Martelli:2023oqk}.

First, we may replace $B_4$ by a complex toric orbifold. 
Here the fixed points $x_a$ of the $U(1)^2$ action 
are now orbifold points, with tangent space 
$\C^2/\Z_{d_a}$, where more generally $d_a=\det (v^a,v^{a+1})$. 
The latter is positive, with appropriately oriented cyclic ordering of the toric divisors, 
and is equal to 1 when $B_4$ is a smooth manifold.  

Appropriate factors of $1/d_a$ then enter fixed point formulae, as also explained in \cite{BenettiGenolini:2023kxp,BenettiGenolini:2023ndb}. In particular, the $S^4_a$ cycles over the fixed points $x_a$ on $B_4$ do not belong to the same homology class, but instead 
the classes $[d_a S^4_a]\in H_4(M_8)$ are all equal. 
Correspondingly the fluxes through the cycles are not 
 \eqref{NS4} but instead
\begin{equation}
N_{S^4_a} = - \frac{1}{8\pi\ell_p^3} \frac{1}{d_a} \frac{y^a_N - y^a_S}{b_1^a b_2^a} \, .
\end{equation}
However, from the above remarks the combinations $d_a N_{S^4_a}$ are necessarily all equal, and we label these by $N_{S^4}$, so that \eqref{eq:ya} is not modified. The same orbifold order appears in the BVAB formula used to compute \eqref{centralchge1}, which generalizes to
\begin{equation}
	c = \sum_{a=1}^d \frac{1}{d_a} \frac{1}{\epsilon^a_1 \epsilon^a_2} (b_a^1 b^2_a)^2 (- N_{S^4})^3 \, . 
\end{equation}
This result provides a gravitational derivation of the gravitational block conjecture 
for M5-branes wrapped on toric four-orbifolds in \cite{Faedo:2022rqx}.
Similarly, the expressions \eqref{epsilon12} for the weights receive a factor of $1/d_a$, whereas \eqref{bi} is formally unchanged.  The expression \eqref{eq:Nai}  for the flux $N^a_i$  is also formally unchanged, but if $B_4$ is not smooth the intersection matrix $D_{ab}$ reads
\begin{equation}
	D_{ab} = \begin{cases} 1/d_{a-1} & b = a - 1 \\ - \det( v^{a-1}, v^{a+1})/d_{a-1}d_a & b = a \\ 1 / d_a & b= a+1 \\ 0 & \text{otherwise} \end{cases} \, ,
\end{equation}
with a similar expression for the generalization of \eqref{Dabc} that may be found in 
\cite{Martelli:2023oqk}.

Second, we may relax the Calabi-Yau condition \eqref{CYcond}, and in particular
\eqref{cons}. 
This is motivated by the so-called anti-twist, discovered in \cite{Ferrero:2020laf} as a novel 
way to preserve supersymmetry 
for D3-branes wrapped on a two-dimensional orbifold known as a spindle, 
but which has since been generalized to many other setups. In particular 
the solutions  \cite{Cheung:2022ilc}, describing M5-branes wrapping orbifolds 
$B_4$ that are the total spaces of spindles fibred over 
spindles, have been further studied in \cite{Faedo:2022rqx,  Martelli:2023oqk, Couzens:2022lvg, Bomans:2023ouw}, 
where it was proposed to relax the second condition in 
\eqref{cons} to 
\begin{align}\label{sigma}
\sum_{i=1}^2 \mathfrak{p}_i^a = \sigma^a\, .
\end{align}
Here  $\sigma^a=\pm 1$ is {\it a priori} chosen 
freely for each toric divisor. Again using 
\eqref{bi} one can check that the following identity 
now holds, for each $a=1,\ldots,d$:
\begin{align}
b_1^a + b_2^a + \sigma^a \epsilon^a_1 +  \sigma^{a+1} \epsilon^a_2  = 1 \, .
\end{align}
From the discussion in the appendix, we may interpret this 
as a necessary condition for the spinor $\epsilon$ to have 
R-charge $\tfrac{1}{2}$, but where the projection 
conditions on the spinor at different fixed points now 
depend on the choice of $\sigma^a$. In particular, 
the chirality of the spinor at $x_a^N$ is determined by the sign of
$\sigma^a\sigma^{a+1}$. This change of chirality at different fixed points
is understood in detail for the spindle  \cite{Ferrero:2021etw}, 
and indeed the motivation for 
introducing $\sigma^a$ in \cite{Faedo:2022rqx} was precisely that 
this describes the known supergravity solutions where
$B_4$ is a spindle fibred over another spindle \cite{Cheung:2022ilc}. 

It would be interesting to understand better what global constraints 
there are on the choice of projection conditions in a general setup, 
but we leave this for future work. 
 
\section*{Acknowledgments}
This work was supported in part by STFC grants  ST/T000791/1 and 
ST/T000864/1.
JPG is supported as a Visiting Fellow at the Perimeter Institute. 
PBG is supported in part by the Royal Society Grant RSWF/R3/183010.

\bibliography{helical}{}

\subsection{Chirality, charge relation for Killing spinors}

We consider Killing spinors with definite charge under a Killing vector $\xi$, and show how this is related to the chirality of the spinor at a fixed point. Suppose the charge is $q$, so that
$\mathcal{L}_{\xi}\epsilon 
\equiv \xi^a\nabla_a\epsilon + \frac{1}{8}(\dd\xi^{\flat})_{ab}\gamma^{ab}\epsilon= \ii q \, \epsilon$. Also
suppose the Killing spinor satisfies a Killing spinor equation of the form
$\nabla_a\epsilon=(\mathcal{M}\cdot\gamma)_a\epsilon+(\mathcal{N}\cdot\gamma)_a\epsilon^c$. 

Near a fixed point, we introduce Cartesian coordinates $(x_1, y_1, \dots, x_4, y_4)$ and polar coordinates $(r_i, \phi_i)$ for each plane parametrized by $(x_i, y_i)$. The Killing vector is given by
$\xi = \sum_{i=1}^4 \epsilon_i \, \partial_{\phi_i} = \sum_{i=1}^4 \epsilon_i \left( - y_i \,\partial_{x_i} + x_i \, \partial_{y_i} \right) $.
From the Killing spinor equation we deduce that at a fixed point
$\xi^a\nabla_a\epsilon=0$ and hence 
$\frac{1}{8}(\dd\xi^{\flat})_{ab}\gamma^{ab}\epsilon= \ii q \, \epsilon$, or
\begin{equation}
	\ii q \, \epsilon = \frac{1}{2}\sum_{i=1}^4 \epsilon_i \gamma^{2i-1,2i} \epsilon  \, .
\end{equation}
Thus, $\epsilon$ is an eigenvector of each $\gamma^{2i-1,2i}$, which square to $-1$ and thus have eigenvalues $\pm\ii$. Therefore, a spinor with definite charge $q$ is related to the weights $\epsilon_i$ via
\begin{equation}
	q = \frac{1}{2}\sum_{i=1}^4s_i\epsilon_i \, ,
\end{equation}
for some signs $s_i = \pm 1$. Define the chirality operator by $\gamma_9 \equiv \gamma_{12\cdots 8}$, with
$\gamma_9^2 =  1$ and eigenvalues $\pm 1$. Thus, when acting on a spinor with definite charge $q$ we also have
\begin{equation}
	\gamma_9 \epsilon = \left(\prod_{i=1}^4 s_i \right) \epsilon \, ,
\end{equation}
so the chirality is equal to the product of the signs appearing in the charge.

\end{document}